# Deformation Modes of a Packing of Rigid Grains:

# Rotation, Counter-rotation, dislocation field

## P. Evesque

Lab MSSMat, UMR 8579 CNRS, Ecole Centrale Paris
92295 CHATENAY-MALABRY, France, e-mail evesque@mssmat.ecp.fr

**Abstract:**
*The properties of the processes of deformation of a packing of rigid grains are analysed when the exact distribution of the normal forces at contacts is known. Importance of grain rotation and of counter-rotation of adjacent grains is stressed. It is shown that all deformation modes form an incomplete vectorial space **D**. This allows the statistics of fluctuation to be determined. Parallel with antiferromagnetism is then drawn and frustration of rotation is shown to be the leading dissipating mechanism. However this does not allow to make the parallel with spinglass due to the structure of the space of deformation modes which is vectorial, which ensures response linearity. In order to demonstrate experimentally the validity of the approach and the existence of the incomplete vectorial space of deformation, experimental examples are studied using regular arrays. They do demonstrate the mechanism of counter-rotation and the existence of the vectorial space of deformation. This theory is applied to compare different experiment and simulation results on the deformation of square and triangular lattices of rods, which can only be done through a statistical approach based on existence of different modes of deformation. Non linear behaviour of the modes is demonstrated experimentally; link with the continuous theory of dislocation is made.*

**Pacs # :** 5.40 ; 45.70 ; 62.20 ; 83.70.Fn
______________________________________________________________________

Nowadays, experiments [1] and computer simulations [2-4] on grain assembly have made so much improvement that it becomes possible to compare their results in detail. However, as granular matter is a complex medium, it might occur that such a comparison be not simple: increasing the number of grains increases most likely the number of possible deformation processes and make the mechanics of the packing more sensitive to slight change of initial steric conditions, so that the precise deformation of two equivalent packings under the same load may look quite different. This makes the problem of comparing two experiments or an experiment and a simulation quite intricate. One way to circumvent this difficulty is to treat the deformation process as a statistical mechanics problem; however this requires in turn to settle this mechanics in term of statistical mechanics; it is the aim of this paper; it focuses only on the case of quasi-static deformation.

The main characteristics of the mechanics of granular matter is that it concerns an assembly of grains in contact. These grains deform very little but the packing can deform largely depending on the loading path. This seems to be obvious; however it has few consequences: first of all, the notion of contact is quite important; but the way to define these contacts does not consist in defining only their position; one needs also to know the forces applied there. Secondly, it is also important to know how contacts move and slide during the deformation. This second point makes the mechanics of granular matter quite different from other mechanics at first sight, since sliding is not allowed in the continuum mechanics approach because it generates discontinuity of matter. In the case of granular matter however, contact sliding, contact generation and contact destruction are the main process of deformation; furthermore, contact sliding is the main (or better the only) process of dissipation so that it cannot be forgotten even for the first step of modelling. On the contrary, as it is assumed in general that grains deform only slightly because the contact forces are small enough, one may expect that the approximation which assumes the grains rigid (i.e. hard-grain approximation for physicists) to be appropriate.

It is difficult to know the importance of this hypothesis about rigid grains. In order to evaluate it, let us consider first the following simple example of an inclined bar supported by a vertical wall and the ground, both contacts exhibiting solid friction; this is a well known problem whose solution is as follows: if the bar, the ground and the wall are rigid, calculation shows that the contacts forces are undetermined if the bar is far from sliding; however, forces are known as soon as the system is at limit of sliding. Furthermore, as soon as the bar is elastic, its deformation allows to force the knowledge of the contact forces. So, in other words, as soon as the system can deform (i.e. the bar can either slide or deform) the contact forces are known.

The profound reason for the non uniqueness of the solution in the case of rigid systems is linked to hyperstaticity [5], i.e. to the number of forces which are unknown compared to the number of equilibrium equations: for instance, in the case of the bar, when it cannot move the number of unknown is 4 (in 2d) and the number of equilibrium equations is 3 (one for momentum, 2 for equilibrium of forces); so the problem is undetermined. On the contrary, when sliding occurs, this adds two relations which raised the indeterminacy.

Granular matter exhibits the same kind of problem and similar set of solutions can be obtained for these media: let us assume the grains to be rigid, and let us first consider an assembly which cannot be deformed; the exact contact force distribution is undetermined because the number of unknown forces is too large compared to the number of equilibrium equations; however, each time sliding can occur, this introduces a new relation between normal and tangent forces, which reduces the number of undetermined forces; so, instead of being hyperstatic, the system falls into isostaticity or it is unstable; anyway, it is deformable.

Let us now assume that one knows the distribution of normal forces which will be satisfied when the system will be sliding. This allows to know the tangent forces at sliding contacts; in this case hyper- and isostaticity can be defined as follows: the system will be hyperstatic [5] (i.e. non deformable) if the system dissipates more energy than the external forces generate during the increment of deformation and it will be at the limit of static equilibrium when both works will be equal; it will be unstable in the other case. This is an other way to analyse the stability of a granular material. However, this method lies on the knowledge of the distribution of normal force. It requires then that this normal force is unique, which is not so obvious in the case of hyperstatic systems for which the structure and the grains are both rigid and for which indeterminacy of local forces is a constraint. Nevertheless, this will be assumed in the rest of the paper.

So, this paper tries and investigates the complexity of the deformation process of a granular assembly submitted to a given stress field; it considers only systems which can deform, for which the normal force at each contact will be supposed to be known. This is probably a crude assumption since it is probably not guaranteed in practice because the force distribution which ensures the assembly to be at the limit of equilibrium within definite boundary conditions is likely not unique so that the normal force at a given location could take few discrete values.

The paper is built as follows: the first part describes the theoretical approach under the assumption that the distribution of normal forces at contact points is known; work during deformation is calculated and used to show that distribution of local slidings appear to be a consequence of an optimisation problem; this allows to define a deformation mode, which corresponds to a set of local slidings. It is then shown that the space of all possible modes of deformation, which correspond to a single set of normal forces, defines an incomplete vectorial space. Consequence about the fluctuations will be drawn from the existence of this space.

Then, in the second part, the role of grain rotations is enhanced; this allows to draw a parallel with antiferromagnetic systems; experimental evidence of the efficiency of this analogy is reported. Discussion pursues with the problem of frustration of rotation, which ensures the system to dissipate energy. As soon as the concepts of frustration and of antiferromagnetism are pronounced, one expects the concepts of spinglass [6] to merge; however, it will be shown that this analogy cannot be drawn in the present case since spinglass formalism requires strong non linear behaviour and since the grain interaction exhibits a linear behaviour when the normal forces are known (this is a consequence of the structure of vectorial space. This means that the analogy with spin glass, if it is correct, can only be a consequence of the multiplicity of the distribution of normal forces. However, as it will be demonstrated in part 3, mode evolution can be non linear, which modifies this view point, which may be too simple in many cases.

So, the third part is devoted to compare different experiment and simulations results together, to exemplify the vectorial character of the deformation processes at micro- scale. It will be emphasised that comparison between different results can only be ensured in a statistical way; definition of a basis of deformation modes will be given in the case of a square lattice of rods which are deforming. It will be shown also that direction of slidings is not the direction of deformation; this will help discussing the validity of the demonstration by Rowe of his dilatancy law. The modelling will be discussed in terms of the continuous theory of dislocation, but it will be shown that this approach cannot be as powerful as in the case of elastic materials because it is difficult to relate the density of dislocation to the evolution of the stress tensor.

## 1. Theoretical background

### 1.a Recalls

One can describe the problem of the mechanical deformation of a packing of rigid grains with friction by applying the theorem of virtual works to determine the limit of stability and the different ways the system may start deforming [7,8]. Let **F**(n) the force applied to grain n by the boundary and **f**(n,m) the force applied to grain n by grain m. The virtual work δW may be written as:

$$-\delta W = 1/2 \{\Sigma_{nm} \mathbf{f}(n,m).\delta \mathbf{u}(n,m)\} - \Sigma_n \mathbf{F}(n).\delta \mathbf{u}(n,ext) \quad (1)$$

where the summation runs over the grains in contact or in contact with the boundaries; k is the solid friction between the grains. δ**u**(n,m) (or δ**u**(n,ext)) is any possible (virtual or not) displacement of the point pertaining to grain n in contact with grain m (or with the boundary). Let **N** the force component of **f** perpendicular to the contact surface. So, Eq. (1) can be written:

$$-\delta W = 1/2 \{\Sigma_{nm} k \, \mathbf{N}(n,m) \|\delta \mathbf{u}(n,m)\|\} - \Sigma_n \mathbf{F}(n).\delta \mathbf{u}(n,ext) \quad (2)$$

where ||**v**|| stands for the norm of vector **v**. If we assume that the grains in contact remain in contact during a small deformation, any δ**u**(n,m) may be decomposed into the sum of three rotations, two (δ**Ω**$_n$ and δ**Ω**$_m$) around the two centres of grains m and n, plus one (δ**Ω**$_{nm}$ around the centre of any of the two grains which accounts for a global rotation of the pair: δ**u**(n,m)= δ**Ω**$_n$\***r**(n,m) - δ**Ω**$_m$\***r**(m,n) + δ**Ω**$_{nm}$\*[**r**(n,m)-**r**(m,n)], where the **r**(n,m) are the vectors originating at the centre of grain n and ending at the contact point with grain m; the stars stand for the vectorial product. So Eq. (1) leads to:

$$-\delta W = 1/2 \{\Sigma_{nm} k \, \mathbf{N}(n,m) \, \|\delta\mathbf{\Omega}_n * \mathbf{r}(n,m) - \delta\mathbf{\Omega}_m * \mathbf{r}(m,n) + \delta\mathbf{\Omega}_{nm} * [\mathbf{r}(n,m) - \mathbf{r}(m,n)]\|\}$$

$$- \Sigma_n \mathbf{F}(n) \, \delta\mathbf{u}(n,ext) \quad (3)$$

This way of writing the grain motion allows to keep each pair of grains n and m in contact two by two all along the deformation; however, it does not allows to keep a quadrilateral structure of grains in contact; in order to do so, one has to add to Eq. (3) a set of equations relating the δ**Ω**$_n$ , δ**Ω**$_{n'}$, δ**Ω**$_{nm}$, δ**Ω**$_{n'm'}$ together. Nevertheless, a first step of approximation consists in neglecting this point and to write the evolution as governed by Eq. (3) only.

We turn now to the determination of the deformation process: according to the virtual work theorem [7,8], the sample is stable as far as Eq. (3) is positive whatever the local displacements δ**Ω**$_m$, δ**Ω**$_{nm}$ and δ**u**(n,ext) are; when Eq. (3) becomes 0 for a set [δ**Ω**$_m$, δ**Ω**$_{nm}$,..., δ**u**(n,ext),...] (or for a series of sets), the sample is at the limit of stability and will yield if one increases the stresses; in such a case the yielding process is defined by one of the sets of yielding modes [δ**Ω**$_m$,..., δ**Ω**$_{nm}$, . . . ,δ**u**(n,ext),...] which cancel Eq. (3).

## *1.b Modes of deformation V*

Let us now come back to the previous formulation (Eq. (2)). For later convenience, we associate a single collective index ($\alpha$) to every contact and we consider the relative sliding $\delta\mathbf{u}_\alpha$. Since the distribution of normal forces within the sandpile is fixed, Eq. (2) is:

$$-\delta W = \sum_\alpha k\, \mathbf{N}\, \delta\mathbf{u}_\alpha - \sum_n \mathbf{F}(n).\delta\mathbf{u}(n,ext) \quad (5)$$

Then the total virtual work can be written simply:

$$\delta W = -\delta W_{int} + \delta W_{ext} \quad (6)$$

with $\delta W_{ext} = \sum_{nm} \mathbf{F}(n).\delta\mathbf{u}(n,ext)$. At the limit of stability, the minimum of $\delta W$ is 0 and the deformation is obtained by minimising $\delta W$; this deformation process is characterised completely by the whole set $\{...,\delta\mathbf{u}_\alpha,...\}$. This set $\{...,\delta\mathbf{u}_\alpha,...\}$ may be viewed as a $3N_c$-component vector, where $N_c$ is the number of contacts. Let us assume the macroscopic deformation of the pile is controlled by a single macroscopic degree of freedom (i.e. the variation $\delta h$ of height h, for example), we may parameterise the relative sliding vectors $\delta u$ by the macro-deformation variable $\delta h$:

$$\delta\mathbf{u}_\alpha = (\partial\mathbf{u}_\alpha/\partial h)\delta h = \mathbf{V}_\alpha \delta h \quad (7)$$

This defines a $3N_c$-component vector $\mathbf{V}$, which is the rate of virtual sliding with respect to the parameter $\delta h$. We will call $\mathbf{V}$ a mode of deformation, since $\mathbf{V}$ characterises completely the way the grains move during the deformation. $\mathbf{V}$ optimises the total virtual work, so that $\partial W/\partial h = 0$. According to Eqs. (2), (6) and (7), one has:

$$\delta h \sum_\alpha k\, N_\alpha\, \|\mathbf{V}_\alpha\| = \delta W_{ext} \quad (8)$$

♣ It is worth stressing firstly that in a triaxial test characterised by an overload $q=\sigma_1-\sigma_2$ and a lateral pressure $\sigma_2=\sigma_3$, one may write $\delta W_{ext}/\delta h = q - K\sigma_2 + \delta W_f$ where $\delta W_f$ is the friction work against walls and with $-K = \delta v/\delta h$ is the dilatancy, where v is the specific volume. Note that $\delta W_f$ can be written in a way similar to the other sliding so that the index $\alpha$ can be expanded to span also over the contacts between the grains and the walls. Within this new convention for Eqs. (5) & (7), the left hand part of Eq. (8) includes friction with the walls and $\delta W_{ext}/\delta h = q + K\sigma_2$. In this case, only the left hand part of this equation is unknown from an experimental point of view.

♣ *2$^{nd}$ remark*: though a mode of deformation characterises the evolution of the system, Eq.(8) is not an equation of evolution of the modes of deformation. Such an equation of evolution should involve the derivative of the modes with respect to h and it should be irreversible. Indeed, non-linearity of the friction force manifests itself in the fact that under the same mechanical constraints if $\mathbf{V}$ is a mode of deformation, then $-\mathbf{V}$ is not a mode of deformation in general (i.e. except in some bifurcation case, an example of which is found in the square lattice of Fig. 2). Anyhow, Eqs. (7) and (8) are direct relations linking the deformation mode and the global deformation $\delta h$ and we choose all the components of the modes to be positive definite. Remark also that a mode of deformation is not any sliding satisfying (8), but rather those sliding sets allowed by the geometrical constraints in the packing. In order to characterise the modes of deformation two questions are to be examined in detail: relation between the modes and interactions between the geometry and the modes.

## 1.c  Space D of deformation modes V

Let us assume as in [10] that under some circumstances, the deformation can occur along two different modes of deformation **V**(l) and **V**(2) so that {**δu**$_\alpha$(l)}=**V**(l) δh and {**δu**$_\alpha$(2)}= **V**(2) δh are two sets of possible sliding corresponding to the same macroscopic deformations δh. We demonstrate now that any combination **V**= β**V**(l)+ γ**V**(2) with positive coefficient of **V**(l) and **V**(2) is also a possible mode of deformation: first of all **V**(l) δh and **V**(2) δh optimises Eq. (2). So, as :

$$\|\beta \, \delta \mathbf{u}_\alpha(l) + \gamma \, \delta \mathbf{u}_\alpha(2)\| = \delta h \, \|\beta \, \mathbf{V}(1) + \gamma \, \mathbf{V}(2)\| <$$
$$< \delta h \{\beta \, \|\mathbf{V}(1)\| + \gamma \, \|\mathbf{V}(2)\|\} = \beta \|\delta \mathbf{u}_\alpha(l)\| + \gamma \|\delta \mathbf{u}_\alpha(2)\| \qquad (9)$$

So, {β**δu**$_\alpha$(l)+γ**δu**$_\alpha$(2}= δh{β ||**V**(1)|| + γ ||**V**(2)|| shall also optimise Eq. (2) if **V**(1) and **V**(2) optimises Eq. (2).

Thus, if **V**(l) and **V**(2) pertain to the space of deformation mode ***D*** the linear combination β **V**(l) + γ **V**(2), with positive β and γ pertains also to this space ***D***. It means that ***D*** has all the features of a vectorial space except that it is not complete (modes -**V** are not defined in general). Hence, a basis of proper modes will determine the space of deformations and the problem is reduced to the knowledge of these proper modes.

One may then ask whether two different directions of sliding may coexist at a same contact point. The answer is no if we take into account elasticity, since inclined contact forces generate hysteresis in this case [9]. But in the case of perfectly rigid grains, hysteresis would not exist so that a change of sliding orientation would not dissipate energy, when possible; so in the case of rigid grain, such event are possible and further investigation is required. This will be done in the last section.

Anyhow, it seems unlikely that this kind of situation can occur when the total number of contacts where two directions of sliding are possible at the same time is about the same as the total number of grains of the sample since the distribution of sliding contacts will span over the whole pile homogeneously in this case, which implies a homogeneous deformation of the pile in two different ways, which means in turn very unusual conditions of grain imbrication. So, one shall expects more likely that cases where different modes of deformation occur at the same time is obtained when these modes are localised in different part of the sample or span over different groups of contacts. This will be exemplified in section 3.

Indeed, this localisation of the deformation modes is observed in the case of the triangular lattice of Fig. 1 for instance.  It is also observed in Fig. 2 in the case of square-lattice samples where a possible mode consists in any pair of two adjacent horizontal rows in which the disks of the upper row rolls in one way and those of the lower one roll in the other way. In section 2 and 3, this mode of deformation will be called a counter-rotating mode. In this peculiar case of Fig. 2, the sliding contacts are the contacts between grains in the same horizontal row, and the grains in two different rows are rotating without sliding; applying the above theory lets expect that any combination of two deformation modes will be possible at the same time; this is indeed what it is observed. When different modes coexist, the effectively observed mode is one of the possible combination and its statistics of real achievement should reflect the statistics of random combinatory see section 3 for further discussion, and for limitation). Furthermore, as the symmetry of the sample imposes that the motion of the rows can occur in both the +x and the -x ways, the space of deformation in this peculiar case of Fig. 2 is a complete vectorial space.

### *1.d Space S of transformation modes* **T**

From Eq. (8) , one can also define a transformation vector **T** and a space *S* of transformation as follows:

$$\mathbf{T}= \{…,T_\alpha,…\}= \{…, k\, N_\alpha\, \mathbf{V}_\alpha ,…\} \quad (10)$$

With the norm of T given by

$$\|\mathbf{T}\| = \Sigma_\alpha\, k\, N_\alpha\, \|\mathbf{V}_\alpha\| = \delta W/\delta h \quad (11)$$

In the same way, if two transformations $\mathbf{T_1}$ and $\mathbf{T_2}$ are possible, any combination $a_1\mathbf{T_1}+a_2\mathbf{T_2}$ , with $a_1+a_2=1$ , $a_1>0$, $a_2>0$, is possible, provided that the set $\{…,N_\alpha,…\}$ are the same for $\mathbf{T_1}$ and $\mathbf{T_2}$ and that the mode evolution is linear; so the ensemble of transformation form an incomplete vectorial space *S* [10]. This space is called the space of transformations .

### *1.e Statistics on the space of transformations*

We have shown that a space of solutions corresponds to the optimisation problem of deformation of a sandpile. This space has the structure of an incomplete vectorial space if the compounds of the force normal to the contacts are fixed; this is what we have always assumed. Hence, the only analytical procedure to determine the effective transformation of the system is to use statistical mechanics methods. We define a microcanonical ensemble of transformations (or deformations) by all samples with the same $\delta W_{ext}$ , the same topology and the same set $\{…,N_\alpha,…\}$.

The points of the space of modes are associated to independent samples which have evolved independently. Suppose that there are n independent proper modes of transformation. Since the space *D* (or *S*)is spanned by these modes, the general mode of deformation (transformation) is a random vector with n random components. We admit also that because of hardening there is a cut-off on the modulus of the modes so that the modes evolve in a confined space.

On the other hand, the macro-transformation is related to the modulus $\|\mathbf{T}\|$ of the mode **T** which is by definition a sum of its components (see Eq.l0). It follows that according to the central limit theorem $\|\mathbf{T}\|$ has a Gaussian distribution with a positive mean value $\|\mathbf{T}\|$ and since there are n free degrees of freedom, its square mean fluctuation $\Delta\|\mathbf{T}\|$ is given by

$$\Delta\|\mathbf{T}\|/<\|\mathbf{T}\|> \sim n^{-1/2} \tag{12}$$

Since $W_{ext}$ is constant, and using the canonical space, one gets from Eq. (12) also an order of magnitude of the fluctuations of $\delta h$ and K:

$$\Delta|\delta h|/<|\delta h|> = \Delta|\delta K|/<|\delta K|> \sim n^{-1/2} \tag{13}$$

These fluctuations have to be interpreted as the fundamental non-reproducibility of experiments on a sandpile. The total number of modes should be much smaller than that of contacts or grains. However, the number of modes is expected to increase with the number of grains. So macroscopic sandpiles contain a great number of modes and have a well-defined predictable average behaviour.

A singular consequence of Eq. (13) is that the more one tries to control experimentally the transformation process (by surveying the boundary conditions, for example), the less will be the number of modes and so the larger are the fluctuations from sample to sample. However, it might occur that the real cause of fluctuations is the non linear character of the evolution of the deformation- (transformation-) modes when macroscopic deformation proceeds, so that the validity domain of Eq.(13) may be restricted to infinitely small deformation.

## 2. Parallel with disordered antiferromagnetism problem of classical spins [6]

### 2.a Parallel with classical antiferromagnetism

Let us now restart from Eq. (3). This equation gives the work due to the packing deformation in terms of local rotations of grains $\delta\Omega_n$ and of the local rotation of pairs of grains in contact $\delta\Omega_{nm}$. An additional simplification of Eq. (3) which is discussed in [7,8] consists in taking approximately into account the $\delta u_{nm}$ terms by modifying the $\mathbf{r}(n,m)$ distances to $\mathbf{r'}(n,m)$ and writes:

$$-\delta W = 1/2\{\Sigma_{nm}\, k\, N(n,m)\|\delta\Omega_n * \mathbf{r'}(n,m) - \delta\Omega_m * \mathbf{r'}(m,n)\|\} - \Sigma_n \mathbf{F}(n).\delta\mathbf{u}(n,ext) \tag{14}$$

Let us now assume, as we did already, that each solution of Eq.(14) can be parameterised with a unique infinitely small $\delta h$ parameter and write that each set $\{...,\delta\Omega_j,...\}$ of solution of Eq. (14) can be written as $\{...,\delta\Omega_i=\Omega_i\, dh,...\}$. Secondly, we write the Hamiltonian $\boldsymbol{H}$ of a magnetic material as $\boldsymbol{H} = 1/2\, \Sigma_{nm}\, J_{nm}\mathbf{S}_n\mathbf{S}_m = 1/2\, \Sigma_{nm}\, J_{nm}\{(\mathbf{S}_n+\mathbf{S}_m)^2 - (\mathbf{S}_n^2 + \mathbf{S}_m^2)\}$. In quantum mechanics, the operator $\mathbf{S}_n^2$ has a constant and unique eigen value. So, $\boldsymbol{H}$ may be rewritten as $\boldsymbol{H}_1 + \boldsymbol{H}_o$, with $\boldsymbol{H}_1 = 1/2\, \Sigma_{nm}\, J_{nm}(\mathbf{S}_n+\mathbf{S}_m)^2$, with $\boldsymbol{H}_o = \Sigma_{nm}\, J_{nm}\mathbf{S}_o^2 =$ constant for a set of coupling strength $J_{nm}$. Then, the first term of Eq. (14) looks much like $\boldsymbol{H} + \boldsymbol{H}_o$; the only difference arises from the way the vector lengths are computed: it is with the use of absolute values $\|\|$ in Eq. (14), but by squaring the quantum scalar-product operator in $\boldsymbol{H} + \boldsymbol{H}_o$. Nevertheless, there is a strong analogy between $\boldsymbol{H} + \boldsymbol{H}_o$ and Eq. (14) if one transposes as follows:

$J_{nm} \leftrightarrow k \, \mathbf{N}(n,m)$ and $\mathbf{S}_n \leftrightarrow \mathbf{\Omega}_n$.

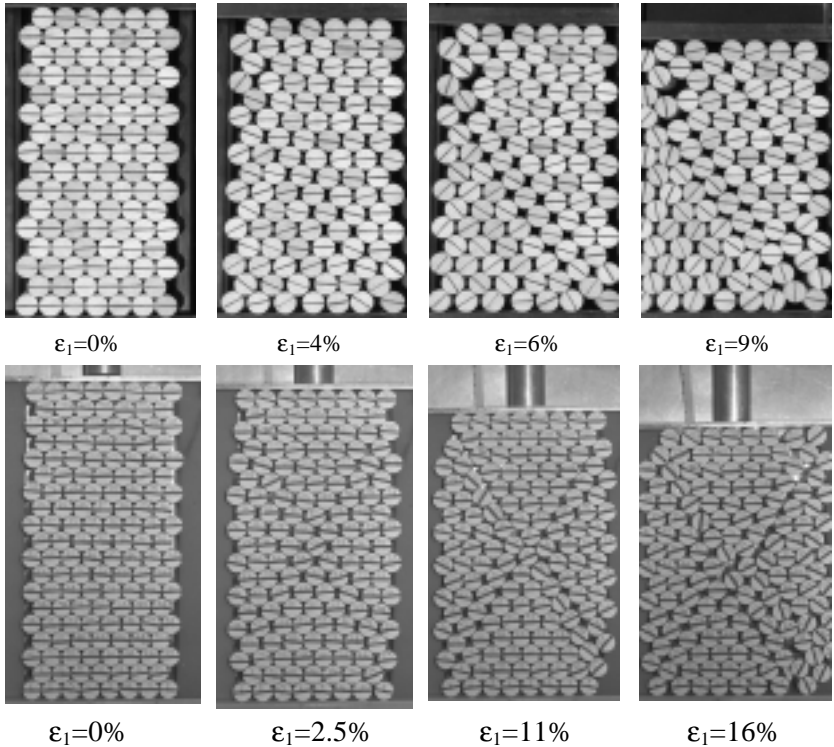

**Figure 1:** *2 examples of the deformation of a* 2d *ordered triangular lattice of rods.*
  **Top-row figures:** *lateral walls are rigid and allows rotation of the grains in contact with boundaries. The first stage of deformation* (4%) *consists mainly in a dilation which allows to reduce the number of contacts/grain to 4. Then, one can observes that deformation occurs along a localised zone inclined at 30° from vertical. This zone of 1 rod thickness separates the medium into two distinct zones. The motion of these zones is possible due to a rolling of the grains at the interface. This set of rolling forces the grains which are adjacent to rotate in opposite direction. So, in each zones two adjacent grains in contact are counter-rotating to minimise dissipation in the bulk and dissipation with walls (after W. Mezftah).*
  **Bottom row Figures:** *This boundary condition imposes a no-slip boundary condition so that counter-rotation process of Top Fig. is mainly forbidden here, except for the lateral part after $\varepsilon_1=10\%$. So, boundary conditions play an important role.*

Furthermore, it turns out that the dimension of the interaction is important in the problem of antiferromagnetism; this is linked to the nature of the $J_{nm}$ coupling which can be a complete matrix or only a projection matrix. And one shall distinguish between the Ising-like, the Heisenberg-like or the 3d kinds of antiferromagnetisms. In the Ising-like antiferromagnetism, there is a unique direction of interaction, i.e. z, and the Hamiltonian writes: ***H*** = $1/2 \, \Sigma_{nm} \, J_{nm} \mathbf{S}_n \mathbf{S}_m = 1/2 \, \Sigma_{nm} \, J_{nm} \mathbf{S}_{zn} \mathbf{S}_{zm}$; i.e. only the projection of spin in a precise direction z is important in the Ising case. In the Heisenberg-like antiferromagnetism, the spins interact in two directions, i.e. x,y , and the $S_x$ and $S_y$ compounds are coupled so that ***H*** = writes: ***H*** = 1/2 $\Sigma_{nm}$ $J_{nm}S_nS_m$= 1/2 $\Sigma_{nm}$ $J_{nm}${$S_{xn}S_{xm}$+ $S_{yn}S_{ym}$ }. At last, the most general 3d case was described in the previous paragraph.

When the study of the deformation is limited to the case of a packing of parallel rods, the problem is obviously 2d and the rotation of the rods can occur in a single direction perpendicular to the packing and parallel to the rod axis. So, in this case, the analogy which shall be drawn is the one with the Ising antiferromagnetism. The efficiency of this comparison is obvious even at first glance to Figs. (1) and (2).

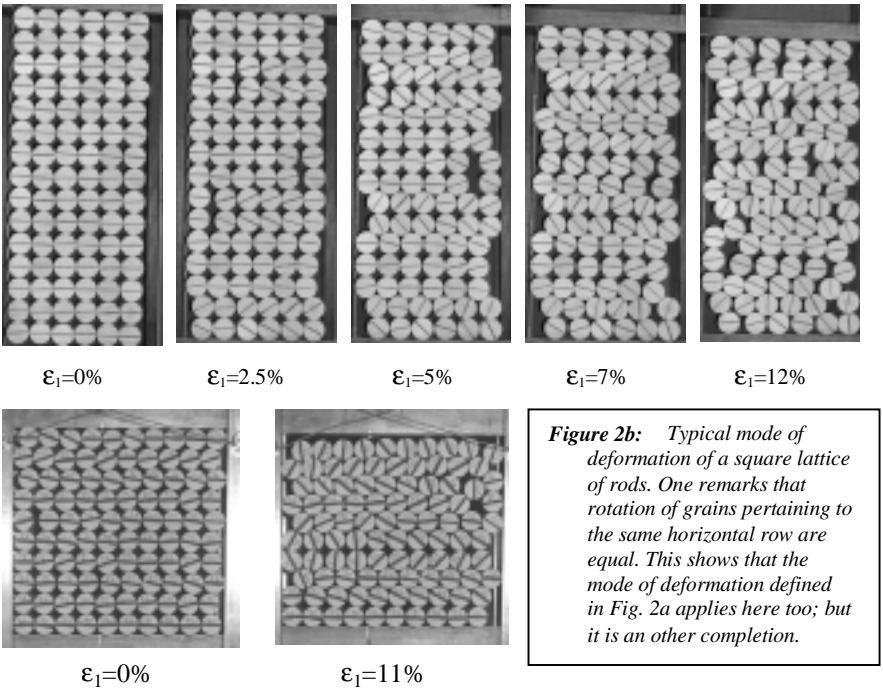

$\epsilon_1$=0%   $\epsilon_1$=2.5%   $\epsilon_1$=5%   $\epsilon_1$=7%   $\epsilon_1$=12%

$\epsilon_1$=0%   $\epsilon_1$=11%

*Figure 2b:* *Typical mode of deformation of a square lattice of rods. One remarks that rotation of grains pertaining to the same horizontal row are equal. This shows that the mode of deformation defined in Fig. 2a applies here too; but it is an other completion.*

***Figure 2:*** *2 examples of the deformation of a 2d ordered square lattice of rods.*
  *Fig. 2a: Top-row figures: the deformation occurs by the motion of pairs of horizontal lines of rods. Top Figures: At $\epsilon_1$=2.5%, the single pair of bottom lines has moved towards the right. At $\epsilon_1$=5% ,the 6$^{th}$ & 7$^{th}$ rows have also moved, and the 12$^{th}$ & 13$^{th}$ has started moving. At 7%, the 9$^{th}$ & 10$^{th}$ rows have also moved. Motion of rods occurs without sliding so that grains in contact are counter-rotating (after W. Meftah).*

In Fig. (1) for instance, one sees two zones separated by a diagonal array of grains where deformation is localised; each zone is made of rods in a lozenge structure which allows a mode of counter-rotation in which each grain in contact turn in a direction and every other grains in the opposite direction to minimise the dissipation. Changing the boundary condition changes the amplitude of the rotations (Fig. 1b).

In the same way, the deformation which appears in Fig. 2 is such as the adjacent grains from two next horizontal rows are counter-rotating (i.e. rotating in opposite direction), whereas grains in the same horizontal row rotate in the same direction. So, sliding occurs only between grains pertaining to the same horizontal row.

## *2.b  Effect of frustration*

What is frustration: consider 3 grains in contact; imposing the rotation of one grain in a direction forces the two others to rotate in the opposite direction; but if these two other grains are in contact, this contact is sliding and dissipates energy. In such a case, one says that this set of three grains are frustrated because at least one of the contact is sliding. This notion can be extended to any loop made of an odd number of grains in contact; on the contrary, any loop with an even number of grains is not frustrated.

This notion of frustration is efficient because the k $N_{nm}$ are always positive quantities. It should be modified if k $N_{nm}$ could be negative; anyhow this is not possible in the case of solid friction. But it is because the k $N_{nm}$ are always positive quantities that one can draw the analogy with an antiferromagnetic Ising model for the deformation of a 2d packing. Furthermore, it can be shown using results on antiferromagnetic systems that disorder of the lattice structure can lead to frustration due to the variation of the coordination number from site to site and is sufficient to generate a complex problem of optimisation. From an experimental point of view, we observe indeed that the interaction imposes quite often a counter-rotation process for the state which dissipates the minimum of energy, as in the case of an anti-ferromagnetic structure where an alternate series of spin-up spin-down state is observed. We also observe that the natural packing which allows deformation is made of lozenges instead of triangle; this is to minimise frustration; it is also to reach a state of isostaticity, for which the number of unknown forces shall correspond to the number of stability equation imposed by the different grains.

Anyhow, solving Eqs. (1) or (2) is an optimisation problem which looks like the one of minimising $H_1 + H_o$, that is to say of finding the ground state of a distribution of magnetic spins interacting through local anti-ferromagnetic couplings of variable strength k $N(n,m)$. It is worth noting however that the terms $r'(n,m)$ entering Eq. (14) is not exactly *($r_n-r_m$); as a consequence, the exact solution of Eq. (14) is not that one of the optimisation of the counter-rotation problem in the precise lattice structure which corresponds to the packing. This can be seen easily in Fig. 2 for instance: Optimisation of counter-rotation problem on a square lattice is possible without generation of sliding on a square lattice; on the contrary, optimisation of the exact deformation mode leads to sliding at contacts between grains pertaining to the same horizontal row. So, the visual solution made of counter-rotation is only an approximate solution, which does not take into account the precise deformation process in detail.

An other source of approximation comes from the possible fluctuations of the coupling coefficients k N(n,m) which may vary from point to point; this will lead the real solution to differ from the one which is obtained from simple topological consideration only. This degree of approximation may look rather important because we know from Dantu experiment [11] that the coupling strength **N**(n,m) varies strongly from site to site but are aligned along l-D channels, so that effect of local frustration of rotation is probably less efficient than it could be thought at first sight.

Anyway, one may find some local example of antiferromagnetic-like coupling in Fig. (1) and (2); which demonstrate the efficiency of this approximation.

♣ *Mean rotation shall be approximately 0:* An other point which is worth noting is that the mean rotation is approximately zero in Figs (1) & (2). The antiferromagnetic-like model allows to understand this result as a consequence of energy minimisation; however, it needs crude approximation and better understanding shall require much larger effort. Anyway, we can also conclude from this modelling that if the whole sample rotates around itself the mean rotation per grain shall be the one of the sample; so this simple modelling allows to interpret the experimental finding reported in [12] by Calvetti et al.

## *2.b Towards spinglass or not?*

It is often thought that systems which exhibits antiferromagnetic-like interactions associated with structural disorder and frustration shall exhibit a spinglass behaviour [6]. This is also what we thought at the beginning [7,8,13]. In this subsection we examine this problem in some more detail and show that the system does not exhibit spinglass behaviour if the set of coupling strength k N(n,m) is unique and well defined and if the mode evolution is linear; however, if the system exhibit more than one set of k N(n,m) at the same time, or if the mode evolution is non linear, spinglass behaviour can be recovered.

♣ *What is a spinglass phase:* main characteristics of spinglass arises from non linear interaction; which makes the problem of minimisation to exhibit a set of local minima. It results from this that the solution the system can chose is not the most optimum but only a local optimum. It can be also shown that this set of local minima exhibits in general a hierarchical structure.

♣ *the case of granular assembly with well definite unique set of {…, N(n,m),…}:* in this case the N(n,m) are fixed; so the space ***D*** of all possible deformation forms an incomplete vectorial space (see section 1). It results from this that one can goes from a solution to an other one continuously. So, the optimisation problem of finding the infinitesimal deformation exhibits a single minimum and does not exhibit local minima. This space of deformation modes can not generate spinglass structure. However, it can generate irreversibility and sample-dependent evolution, when the {…, N(n,m),…} set can evolve non linearly after each incremental deformation in an irreversible way; this will be exemplify in section 3.

♣ *What does it occur when the set of {…, N(n,m),…} is not unique:* in this case two cases can be considered; in the first case, the {…, N(n,m),…} set of parameters fixed for a long while; in the second case, the {…, N(n,m),…} set of parameters jumps rapidly from a configuration to an other configuration. In the first case, when the {…,N(n,m),…} are frozen, the different deformation spaces corresponding to each {…,N(n,m),…} configuration are disjoint. In this case, the deformation exhibits a structure of spinglass.

In the second case, the systems jump from one {…,N(n,m),…} configuration to an other one rapidly; in this case it is worth using the transformation formalism described in subsection 1.c ; as it was shown that the space *S* of transformation modes T exhibits also the structure of an incomplete vectorial space, this allows to restate the problem in a simpler manner which does not exhibit spinglass behaviour, as far as the evolution of the modes remain linear, which is not the case in examples given in section 3.

## 3. Comparison with simulation and experimental results

From previous sections, we can sum up the situation in the following way, when the normal forces N(n,m) are known: the space ***D*** of all deformations is an uncomplete vectorial space so that the really observed deformation path can be any one of the combinations of these deformations, with equal probability of achievement. This allows to treat the problem using a statistical treatment and to associate an entropy of deformation to the considered situation.

A consequence of this result is that the deformation of a granular sample is not unique and that comparison between two sets of experiments shall be done in a statistical meaning. Indeed, the non uniqueness of the response is observed in Figs. (1) & (2), where two sets of experiments are reported each time and where these two sets are not identical. We want to show in this section that comparison between these results can be performed; but this requires to define the deformation modes or the transformation ones. Let us look successively to the two cases of Fig. (1) and Fig. (2).

### *3.a Deformation of a triangular lattice*
This deformation is reported in Fig. (1). The deformation starts with the transformation of the triangular structure into a lozenge lattice since the grains in a same row loose contact with adjacent grain. This deformation mode is valid for both samples; this is due to the hyperstatic character of the triangular lattice which is hence not deformable. Then, in both cases of Fig. 1.a & 1.b, the deformation becomes localised on an inclined row of grains when it proceeds whose inclination is 30° compared to vertical. It corresponds to a principal direction of the triangular lattice.

In both cases (Fig. 1.a & 1.b), this inclined row allows the triangular upper zone to roll on the triangular bottom zone so that the dissipation is mainly produced at contacts between adjacent grains pertaining to the inclined row forming the interface.

Further deformation proceeds in the example of Fig. (1.b) ($\varepsilon_1$=11%, 16%); it is achieved through the generation of a new inclined row at 30° from vertical and symmetric from the previous one, where deformation is localised.

It means that the mechanism of localisation of deformation into a single row inclined at 30° degree from vertical is the preferential mode of deformation for this kind of triangular lattice oriented as in Fig. (1). However it can occur at different location (as shown in Fig. 1.a-$\varepsilon_1$=9%).

Nevertheless, grain motion is in fact more complex, since Fig. (1.a) exhibits also two systems of counter-rotation of grains, one in each triangle , i.e. top-right and bottom-left, whose amplitude is more pronounced in the bottom-left. This system of grain rotation is made possible because the contacts have a lozenge structure; it is caused most likely by some motion of the vertical and horizontal boundaries. These sets of combined rotations are less visible in Fig. (1.b), at least when $\varepsilon_1$<4%, for which vertical boundaries were deformable; it occurred only for $\varepsilon_1$>10% (Fig. 1.b) in the two lateral zones in contact with the vertical walls; grain rotation is caused likely there by the relaxation of the elastic deformation of the vertical walls which occurs when the sample is too much deformed; this happens after large deformation only due to the softness of the lateral walls which were made of expanded elastic material.

So, these modes of counter-rotations are independent of the localised deformations and can be combined to them in the proportion imposed by boundary conditions. We believe that this shows the efficiency of the above theory . This effect may be also related to the so-called "soft modes" introduced in [5].

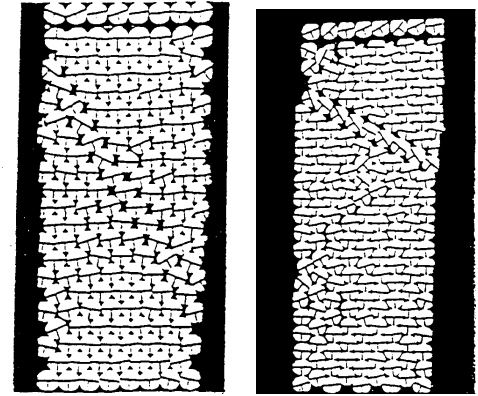

*Figure 3:* *vertical compression of two triangular lattice stuctures of rods with two different orientations. Note the inclination of the localisation band which corresponds to the direction of a principal axis. (after W. Meftah)*

At last, it is worth noting that the notion of mode of deformation shall be preserved when changing the orientation of the principal stress direction compared to the lattice structure. However, its structure can be modified. This is indeed what can be seen from Fig. (3), which reports the typical mode of deformation of two triangular lattices of rods with different orientation of the lattice and submitted to loading with the same principal stress direction. Indeed, one can observe the same kind of localised motion occurring on a single inclined row of grains. However, the two inclinations are different , i.e. 60° and 30° compared to horizontal, for the two lattice orientations and correspond to the principal direction of each lattice. One observes also equal rotation of the rods pertaining to the interface row. So, in this case the mode of deformation is preserved but its inclination varies. This demonstrates the large influence of the local order.

## 3.b Deformation of a square lattice

Let us now turn to the case of the square lattice structure of Fig. (2). Fig. (4) is a computer simulation of the same deformation process. Obviously, Fig. (2) and Fig. (4) are not identical, so that one cannot conclude at first sight that the simulation gives the same result as experiment. Furthermore, the mode of deformation looks much more complex here than in the case of Fig. (1). However, looking carefully to the grain motion allows to decompose the deformation mechanism into independent modes. A simple definition of the basis of the deformation modes is as follows: Let us consider the bottom and the top part of the sample to translate horizontally compared to each other, and let us consider that the separation between these two parts is made of a single horizontal row of rods; in this case a possible definition of the mode of deformation will be the way all the grains of this single horizontal row move and rotate in order to allow the translation of the two parts of the sample, cf. Fig. (5). Indeed, from Fig. (2) and (4) one sees that these grains rotate equally in the same direction allowing the translation of the upper and lower parts without sliding; motion of grains inside the horizontal row exhibits sliding on the contrary. So, the deformation mode {…, $\delta\mathbf{u}_\alpha$,…} is made of a single series of sliding vectors, which are vertical and have the same amplitude and direction; these vectors are located at the contacts between the grains of the considered row. The contacts between this series of horizontal row and the upper and lower part of the sample are not sliding.

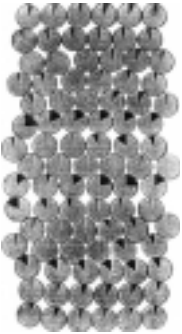

*Figure 4:* Calculation performed with Trubal 2d by Mahboudi & Cambou of the deformationof a square lattice of rods under uniaxial vertical compression.

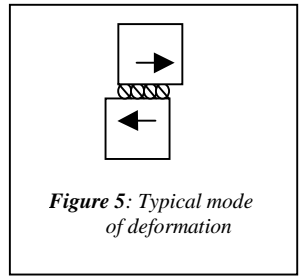

*Figure 5: Typical mode of deformation*

Within this definition of the modes, they are N different modes, if N is the number of horizontal rows forming the sample. It is worth noting that the direction of rolling of the rods can be clockwise or anticlockwise, so that the space of deformation is a **complete** vectorial space (and not an uncomplete one as in section 1). It is also worth mentioning that one can form a new basis of deformation by combining the previous modes together in an adequate manner, so that the definition given above is not unique. In particular one could desire using modes which do not exhibit mean rotation; This can be done by combining the motion of two adjacent horizontal rows, the first one rotating clockwise, the other one anti-clockwise. Both definitions will lead to equivalent descriptions.

Anyhow, it is obvious at first sight that the deformation reported in Figs. (2 & 4) can be decomposed using this basis. This demonstrates that the two results of Fig. (2)

and of Fig. (4) are statistically equivalent; even if they are different. So the **simulation describes correctly the experiments**.

♣ *Remark:* An other point which is worth noting, and which is very peculiar to the square lattice structure, is that the variation of the sample height $\delta h$ scales as $2(r\delta\theta)^2=\|\delta u_\alpha\|^2$, where $\delta\theta$ is the rotation of a grain of radius r. This means that the deformation has (i) a non linear behaviour and (ii) that this non linear behaviour enhances the development of the selected modes. This explains why one observes in Fig. 2 that when a mode is generated it propagates much faster than the others, till complete expansion is reached. Due to the non linear character of the deformation mode, the deformation looks like an irreversible process and like **buckling**.

However, linearity is restored for the combination of fully developed modes of deformation of the same kind. Indeed, this is what is observed on Figs. (2) and (4), for which the deformation progresses by the addition of fully developed modes since $\delta h=\delta h_1+\ldots+\delta h_n$, where $\delta h_i$ is the variation of height of the sample corresponding to the transformation of a line from a square structure into a triangular one. So, within this scheme $\delta h_i$ is quantified and equal to $2r[1-\cos(30°)]$. Within this scheme, the statistics of the deformation can be settled for a given $\delta h \gg 2r[1-\cos(30°)]$. Each mode i, corresponding to the rolling of line i, can be either fully developed or not at all developed. So, considering a large sample made of N horizontal lines and considering a small deformation $n=\delta h/[2r(1-\cos 30)]$, (with $n\gg 0$, but $n\ll N$), the number of possible combinations is approximately $N!/[n!(N-n)!]$. This defines the statistics when $n\ll N$, i.e. when the probability of finding two modes in adjacent position is negligible. For $n\rightarrow N$, this statistics is no more valid.

*3.c non linear behaviour*

It is known from experiment that the deformation of granular samples exhibit memory effect and non linear behaviour. Indeed, existence of soft modes [5] as they are described briefly in section 3.a is a cause of hysteresis and of entropic irreversibility.

Also, examples developed in section 3.b in the case of the deformation of a square lattice have demonstrated the non linear character of the evolution of the modes; this breaks the linearity of the space of deformation, and generates in turn irreversibility since once a mode is selected it develops much faster than the others. But deformation of square lattice is not the only example of non linear behaviour: in the case of the deformation of the triangular lattice, one sees that deformation is localised along inclined rows; indeed, one can see on this figure that the propagation of this mode is also non linear due to the dilatancy mechanism, as shown by Rowe already (see section 3.e below). However, it is less non linear than for square lattice; since this one is strong due to a bad control of the deformation: it would not exist if the width of the sample was used instead of $\delta h$ to control the deformation.

Anyhow, it appears that the non linear character is a common feature which may play an important part in the propagation of the modes. For instance when the propagation of the deformation is related to a decrease of the dilatancy mechanism, one may expect an instable behaviour and the generation of spontaneously fully developed modes of deformation, each one at one time; on the contrary, when the propagation of the mode of deformation is associated to an increase of the dilatancy mechanism the deformation mode will become harder to stimulate so that a homogeneous development of all the modes will be generated at the same time, leading to an homogeneous deformation.

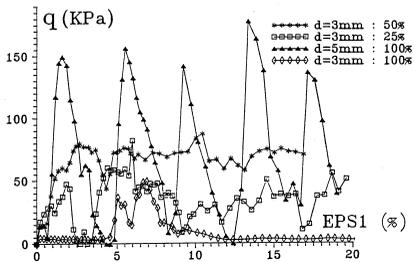

***Figure 6:*** *variations of* $q=(\sigma_1 - \sigma_3)$ *during the compression of 2d packings of rods. The larger variations are obtained for the sample with equal-size rods; periodicity corresponds to the triangular periodic structure.*

Fig. 6 reports the $q=(\sigma_1-\sigma_3)$ variations of few 2d compact samples of rods in a classical 3d triaxial set-up; the curve which exhibits the larger quasi-periodic variations is made of equal rods so that its packing exhibits the triangular lattice structure; the period of variation corresponds to the lattice parameter. This demonstrate the effect of dilatancy.

In other word, non linear nature of the mechanics of the propagation of the deformation mode may modify the proposed theory. This may be more important in 2d than in 3d, since the dilatancy effects in localised bands of deformation are more important in 2d than in 3d for which the notion of critical state and of critical density looks more efficient.

### *3.d link with the continuous theory of dislocation*

To start with, it is worth recalling that the theory of dislocation exists within two schemes [15]; the first one is called the continuous theory of dislocation; it was built to describe fluids; in this theory, dislocations can take any orientation and amplitude.

The second one was built to describe the mechanics of solids and of crystals; it assumes a lattice structure; so that the dislocations can take only a finite set of orientations and discrete amplitudes. This second approach incorporate an elastic term which allows to control the plastic deformation by controlling the motion of dislocations… In this second approach, dislocation was introduced to explain the small limit of plasticity of elastic materials, which cannot be understood if one assumes defects similar to those of Figs. 1 & 3. It results then that these dislocations combine with elasticity destroy these defects; they screen them and confine the lattice anomaly on a single mesh even if elasticity perturbs the lattice structure at larger scale (see Fig. 8).

Obviously, this second approach cannot apply to granular matter. But this will be discussed in more details. So, the aim of this section is to show that only the former approach applies to granular matter. The field of Burgers vector **b** will be defined and found on few examples; it will correspond in fact to the deformation mode **V** defined in section 1. We will show also on few examples that the dislocation field controls the local dissipation but cannot be expressed as a function of the strain tensor.

*Few recalls* [14,15]*:*

♣ *Density of dislocation* $\alpha$: Be **s** the displacement field, one can define the strain tensor as $\beta$=**grad s** . Assuming also that the medium is continuous and remains continuous during the deformation imposes the compatibility equation, i.e. **rot** $\beta$=0. However, if the medium is no more continuous this compatibility equation is no more needed so that one gets **rot** $\beta$=$\alpha$≠0. $\alpha$ is called the density of dislocation.

♣ *Burgers vector* **b**: Consider a closed loop *C* and any surface $\psi$ with incremental element **d**$\psi$ which starts from this loop *C*, and let take the limit of the loop length tending to zero; this allows to define local quantities. In particular, one defines the Burgers vector **b** as the flux of $\alpha$ through this surface . So

$$\mathbf{b} = \int_\psi \alpha \cdot \mathbf{d\psi} \qquad (15)$$

Within this definition, **b** is a field so that it depends on the precise location **r**. One sees that **b** is related to the field of discontinuity of **s**: **b**(**r**)=0 if the displacement field **s** is continuous at location **r** and **b**(**r**) is non zero when discontinuity of **s** exists at point **r**. Further definition and examples of applications can be found in refs. [14,15].

*Physical interpretation of the Burgers vector and definition of the Burgers field:*
Consider a granular material submitted to a deformation $\delta h$; it is obvious that discontinuity of s occurs only at sliding contacts; so, writing Burgers vectors as $\delta\mathbf{b}(\mathbf{r})=\mathbf{b}(\mathbf{r})\,\delta h$ , one gets that **b**(**r**) is zero inside a grain or at any contact between two grains which roll without sliding; it is non zero at contacts n which slides and it corresponds just to the sliding vector $\delta\mathbf{b}(\mathbf{r})=\mathbf{b}(\mathbf{r})\,\delta h= \delta\mathbf{u}_n$. So, one can identify the Burgers field to the deformation mode **V**={…, $\delta\mathbf{u}_n$,…}/$\delta h$ and write the field **b**(**r**):

$$\mathbf{b}(\mathbf{r}) = \{\ldots, \delta\mathbf{u}_n, \ldots\}/\delta h = \mathbf{V} \qquad (16.a)$$

Caution has to be taken when the mode of deformation does not vary linearly with $\delta h$; in this case one shall introduce the macroscopic strain tensor $\varepsilon$ to control the system, where $\varepsilon$ is assumed to derive from a continuous displacement field , i.e. **rot** $\varepsilon$=0, and to fit in average the macroscopic deformation <$\beta$> of the sample. This will lead to define deformation modes $\mathbf{V}_\varepsilon$ and Burgers vector field $\mathbf{b}_\varepsilon(\mathbf{r})$ as depending on $\varepsilon$:

$$\mathbf{b}_\varepsilon(\mathbf{r}) = \{\ldots, \delta\mathbf{u}_n, \ldots\}/\varepsilon = \mathbf{V}_\varepsilon \qquad (16.b)$$

*Strain tensor $\varepsilon$ alone cannot describe the field of discontinuities:*
Indeed, the strain tensor $\varepsilon$ is defined using the continuous medium approximation, so that **rot** $\varepsilon$=0; it has to be differentiated from the true tensor $\beta$ derived from the true displacement field **s**, i.e. $\beta$=**grad s** ; this last one exhibits a non zero rotational, i.e. **rot** $\beta$≠0, and contains all the discontinuities due to grain motion at sliding contacts: hence $\varepsilon$ cannot describe the sliding (i.e. deformation) modes **V**. However, $\varepsilon$ controls the external work $\delta W_{ext}$ of Eq. (6) and it can be used to parameterise the amplitude of the sliding field with the deformation. Indeed, when deformation modes were defined, i.e.

Eq. (7), it was assumed that a single parameter, i.e. δh, was needed to parameterise the deformation process; it turns out that it is not sufficient and that one needs the whole stress tensor **ε** to achieve the entire parameterisation.

We see now how the present theory applies to concrete systems.

*Application to particular examples:*

Figs. (1-4) are examples of sample deformation with samples exhibiting different lattice structures. In all cases, the deformation is ensured by the motion of one or few rows of rods; this (these) rows can be inclined compared to horizontal (in the case of Figs 1,3) or not (Figs. 2,4); but in all cases deformation looks localised on a shear band. It is worth noting now that sliding direction occurs always perpendicularly to the row inclination. So, this demonstrates that the shear band direction is not the direction of Burgers vector **b**. However, this figures seem to indicate also that the shear band is the zone where the dislocation field is important, although I do not know any demonstration of this point. Fig. (7) reports two experiments on passive failure in a systems of rods with a horizontal free surface and, hence, submitted to a non homogeneous stress field. These two results strengthen the above analysis.

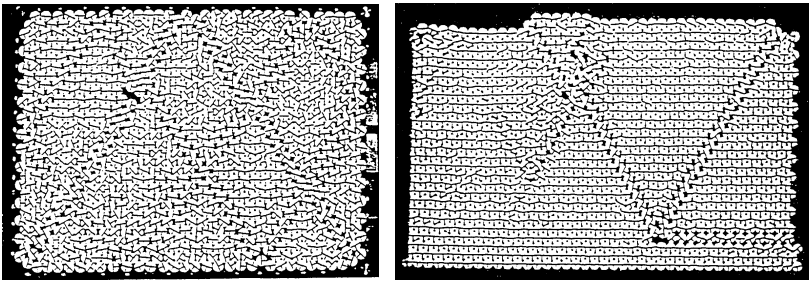

*Figure 7: passive failure obtained in two triangular structures of rods with a horizontal free surface. Deformation are localised along a set of 1 (a) or few (b) inclined rows of rods; sliding occurs mainly between the grains of this (these) row(s) and sliding directions are perpendicular to the row direction(s).*

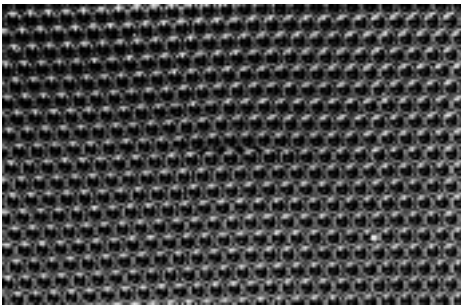

*Figure 8: example of dislocation on a 2d triangular lattice of bubbles. (photo W.M. Lomer, after Bragg & Nye). The topologic defect, which is in the centre, is only observable under skimming view and after a sheet rotation of 30°. This demonstrates that elastic deformation screens the defect. Hence, this pattern is not comparable to those of Figs. 1, 3, 7.*

It is also worth noting that such simple sliding patterns are linked probably to the simplicity of the square or triangular structures used in these examples. However this response is already more complicated than what could expect a priori, since one observes (i) sliding directions which are not the localisation directions and (ii) a set of counter-rotation process which allows to minimise the dissipation of energy, according to the precise boundary conditions. So it is most likely that a more general case, with a more complex internal structure, will not exhibit such simple localisation pattern and such simple distribution of sliding, so that one shall not consider the sliding field to be strictly localised in the "shear zone".

As a conclusion, one shall keep in mind that if the shape of the localisation allows to define the shear zone and the direction of shear, it cannot ensure the determination of the sliding directions nor the one of the dislocation field. This is due simply to the fact that shearing exists already with continuum mechanics and that sliding, dislocation field and Burgers vector do not.

The second point which is worth recalling concern the pseudo "shear banding" which is localised on a single row of grains in simple 2d lattice structures such as the triangular structures of Figs. (1,3,7). These zones are in fact a transition between two compact structure which have the same directions but which have not compatible positions. This row is an adaptive structure which is not analogous to the dislocation mechanism which occurs in elastic materials, because in these materials, the adaptation remains localised on one lattice mesh even if it distort the lattice directions and lengths on further distance; it results from this that it is hard to detect; an example of such a dislocation is shown in Fig. (8) on a bubble network. The elasticity of the atoms or of the grains plays an important role in this mechanism [16] so that dislocation mechanism reduces the limit of plasticity of elastic material; this is not the case for granular materials. Furthermore, the contribution of the disturbance is averaged within one mesh by deformation of the grains in the case of elastic material, since it propagates to infinity in the case of rigid grains. *This is why the row along which the deformation proceeds shall not be considered as a dislocation, but can be better consider* **as a grain boundary**, *which is a zone of transition between two different lattice structures.*

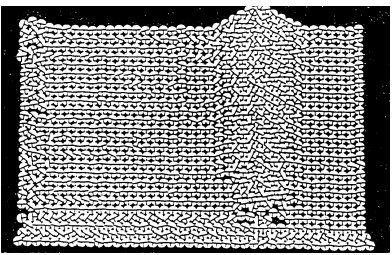

**Figure 9:** *Structure obtained after deforming a square lattice of rods with a top free surface submitted to gravity and to horizontal compression (after Meftah). One sees the condensation of the square lattice into a triangular one; analogy with grain boundary can be drawn.*

Similar analogy with grain boundary can be drawn for the result reported on Fig. (9): here the structure was obtained from the deformation of an initially 2d square lattice structure with a top free surface via a horizontal compression . Indeed, this figure shows the condensation of the square lattice structure into a triangular one. This looks like a stress-induced phase transformation, and the boundary between the two structures appears as the analogue of a **grain boundary** too.

### *3.e Rowe's relation*

The modes of deformation exhibited in Figs. (1-4,7) seem to be good examples of the mechanism assumed by Rowe to find his dilatancy theory [17]: Indeed they exhibit the dilatancy mechanism described by Rowe, which is linked to the motion of the grains inside the deformation zone; they demonstrate also that the average translation of these grains occurs in a direction different from the principal direction of the lattice, and these two directions define the dilatancy. However, this is all what can be compared since the way the system dissipates in both cases are completely different: in the figures, sliding occurs at contacts between grains pertaining to the row where localised deformation develops so that sliding vectors are always perpendicular to the lattice direction; whereas, in the Rowe model [17], the direction of sliding is the mean direction of translation of the grains, since grains are assumed not to rotate. So, it turns out that the system of rotation and of counter-rotation leads to a sliding system which is quite different from the one assumed by Rowe, even if the motion of the grain centres, which characterises the strain tensor $\varepsilon$ are identical in both cases.

Hence, experimental results demonstrate that the theoretical approach proposed by Rowe does not describe correctly the reality, even though the macroscopic experimental law concerning the stress-dilatancy behaviour is in agreement with the relation proposed by Rowe. This is probably a consequence of an even number of errors which compensate themselves two by two.

In turn, the formalism of the continuous dislocation theory can be used to settle this problem in other words. It is known from long that the field of dislocation which correspond to a sample deformation $\varepsilon$ cannot be explicitly described using the field of deformation $\varepsilon$ alone, since these two fields are incompatible, i.e. $\beta \neq \varepsilon$; this is why one cannot calculate explicitly the internal energy dissipated by the sample during the deformation process; however one can calculate the energy furnished by the exterior using the deformation $\varepsilon$. Anyhow, if the system is in equilibrium, each other term shall compensate; this is what Eq. (2) expresses.

To go further, one shall introduce a mechanism based on statistical-physics principles in order to link the two fields (i.e. the field of deformation and the field of dislocation); this will allow to fix in average the production of dislocation for a given stress field according to a given deformation path. Indeed, such a statement does not exists yet for granular matter, even if it exists already for elastic materials and explains why the motion of dislocation depends on the stress field itself in these elastic materials.

### *3.f Last remark: the problem of hyperstaticity*

Let us now consider a 2-D sample as in Figs. 1 and 2 containing $N_o$ grains with $N_c$ contacts; we neglect the boundaries (for a precise discussion see [5]). The number of unknown components is $2N_c$, whereas the mechanical equilibrium leads to $3N_o$ equations and the sliding condition to n more equations ($n<N_c$; $n=N_c$ when each contact is sliding). So, the system is hyperstatic if $2N_c$ is larger than $n+3N_o$; according to the different experimental results (Figs. 1 & 2), the mean coordination number per grain is about $n_c=4$; as any contact pertain to 2 grains, or to a grain and to the boundary, n~$2N_c/N_o$ so that the system shall be hyperstatic if n <$2N_c$- $3N_o$=$N_o$. This means in particular that they are more than one solution for the normal forces **N**(n,m). This hyperstaticity explains also why the method used by Gherby et al. [18] is suitable to determine the distribution of contact forces: if any rod of the system was exactly at the limit of sliding, the mechanical equilibrium would break whatever the pushing force in the transverse direction is.

## 4. Conclusion

This paper shows in section 1 that the problem of deformation of a granular sample is an optimisation problem which leads to define a space of deformation modes which has the structure of an incomplete vectorial space when the set of normal forces at contacts are known and unique. The deformation modes are simple to visualise because they correspond to sliding directions of the contacts. However, this needs to determine the motion and the rotation of each grain. As shown in section 3.d, it turns out that these deformation modes define also the possible Burgers vector field, which defines the field of dislocation. Typical examples were analysed within this scheme.

Furthermore, it was demonstrated in section 1.d that the space of deformation modes is not the most general space which can be used, because it does not work if more than one set of normal forces exist for a given topology of the assembly of grains submitted to a given boundary condition. In such a case, the space of transformation modes turns out to be more efficient. This one has also the structure of an incomplete vectorial space. However, it is less easy to visualise.

Typical examples of the deformation of 2d regular lattice structure of rods were given in section 2 and 3 and their comparison done. Comparison between these different results have successfully been tempted using these spaces and a statistical analysis was defined: in order to compare too results corresponding to the same topology, it is needed first to determine the modes of deformation and/or transformation which corresponds to the structure and to the boundary condition; then to analyse whether the spectrum of proper modes are identical or not. This was performed for triangular and square lattice, and reasonable agreement was found between the different results, so that we could conclude that experiment looks very much the same; it was also used to prove that experiments looks like simulation in the case of square lattice.

At last it is worth strengthening that rotation and counter-rotations of grains play an important in the mechanics of granular media. This is already known [19] but often neglected. For instance, this was neglected by Rowe in his dilatancy model; this is why his approach of the dilatancy-stress law cannot be correct even if the relation he found seem to be verified experimentally. It is also worth recalling: (i) one shall not assimilate a shear zone and a process of localisation of deformation to a dislocation process, both are completely different since in the first case the medium can remain continuous, whereas it never remain continuous in the second case; (ii) topological incompatibility between lattice structure shall not be considered as a dislocation process, but it can be considered as the analogue of a grain boundary in polycrystalline systems; (iii) such a grain boundary cannot exist in elastic materials since the structure will adapt itself spontaneously by forming a dislocation; (iv) anyhow, grain boundary similar to the ones found between elastic poly-crystals can be found in granular matter mechanics when transformation of structure operates (Fig. 9) or when boundary conditions enforces long scale disorder and disinclination of lattice structure. At last, if dislocation combined with elasticity in classical elastic materials allows to adapt the lattice structure to topological perturbation within a mesh, this is not at all what one can see in granular materials; this makes the two mechanics quite different and makes the specificity of granular matter mechanics. On the other hand, granular-matter mechanics is concerned with 3d disordered packing most often; this means that the material (i) contains a large number of topological defects (ii) does not exhibit local order; this reduces most likely the scale of the long range effects described above. However, it is likely important in 2d monodispersed packing.

*Acknowledgements:* This work has benefited from theoretical discussions with F. Radjai at the beginning of his thesis and from experimental results obtained in collaboration with W. Meftah during his thesis. Discussions with D. Bonamy and M. Dubois were also appreciated. CNES is thanked for partial funding.

**References:**

[1] *Powders & Grains 97*, R.P. Behriner & J.T. Jenkins ed., (Balkema, Rotterdam, 1997)
[2] F. Radjai, D. Wolf, S. Roux, M. Jean & J.J. Moreau, "Force network in dense granular media in *Powders & Grains 93*, eds R.P. Behringer &J.T. Jenkins, (Balkema, Rotterdam, 1997);
[3] F. Radjai, D. Wolf, M. Jean & J.J. Moreau, "Bimodal character of stress transmission in granular packing", *Phys. Rev. Lett.* **90**, 61, (1998)
[4] H.J. Herrmann & S. Luding, *Continuum Mech. Therm.* **10**, 189-231 (1998)
[5] A.V. Tkachenko &T.A. Witten, *Phys. Rev. E* **60**, 687, (1999)
[6] K. Binder, A.P. Young, *Rev. Mod. Phys.* **58**, 801-976, (1986); M. Mézard, G. Parisi, M.A. Virasoro, *Spin Glass Theory and Beyond*, World Scientist Lecture Notes in Physics, v o l .9. (1987)
[7] P. Evesque (1993), *Mat. Res. Soc. Symp.Proc.* **291**, 97
[8] P. Evesque & D. Sornette, *J. Mech. Beh. Mat.* **5**, 261, (1994)
[9] R.D. Mindlin and H. Deresiewicz, *J. Appl. Mech*., 327-44, (1953)
[10] F. Radjai & P. Evesque, *Actes du 11ème congrès Français de Mécanique*, Villeneuve d'Asq, 1993, tome IV, p. 469-472
[11] P. Dantu, *Géotechnique* **18**,50-55, (1968); Annales des Ponts et Chaussées IV~ 1-10, (1967)
[12] F. Calvetti, G. Combe & J. Lanier, "Experimental micromechanical analysis of a 2D granular material: relation between structure evolution and load path", *Mechanics of Cohesive-Frictional Materials* **2**, 121-164, (1997)
[13] A. Sornette, D. Sornette, P. Evesque, *Nonlinear Processes in Geophysics* **1**, 209-218, (1994),
[14] F.R.N. Nabarro, *Theory of Crystal Dislocations*, Clarendon press, Oxford, (1967)
[15] D. Bovet, in *Powders & Grains*, pp. 295-302, J. Biarez and R. Gourvès ed., Balkema, Rotterdam, (1989).
[16] C. Kittel, *Introduction à la physique de l'état solide*, Dunod, (1972)
[17] P.W. Rowe, "The stress dilatancy relation for static equilibrium of an assembly of particles in contact", *Proc. Roy. Soc. Lndn* **A269**, 500-527, (1962)
[18] M. Gherbi, R. Gourvès, F. Oudjehane, "Distribution of the contact forces inside a granular material", *Powders and Grains 93*, pp. 167-171, Thornton ed., Balkema, Rotterdam, (1993)
[19] see for instance C.S. Chang & A. Misra, "Modelling of discrete granulates as micropolar continuum", *Journal of Engineering Mechanical Division*, ASCE, 116, 2310-2328 (1990); D. Harris & P. Solberg, "Co-operative motion of grains in discrete model", *Powders and Grains 93*, pp. 87-192, Thornton ed., Balkema, Rotterdam, (1993)
[20] W. Meftah, *Du matériau discontinu formé de grains au milieu continu fictif*, Thèse de Doctorat de l'Ecole centrale Paris, (13 mai 1996)

The electronic arXiv.org version of this paper has been settled during a stay at the Kavli Institute of Theoretical Physics of the University of California at Santa Barbara (KITP-UCSB), in june 2005, supported in part by the National Science Fundation under Grant n° PHY99-07949.

*Poudres & Grains* can be found at :
http://www.mssmat.ecp.fr/rubrique.php3?id_rubrique=402